\documentstyle[emulateapj,danonecolfloat]{article}

\def\beq#1{\begin{equation}\label{#1}}
\def\eeq{\end{equation}}
\def\beqa#1{\begin{eqnarray}\label{#1}}
\def\eeqa{\end{eqnarray}}
\def\eq#1{equation~(\ref{#1})~}

\def\eqq#1{equation~(\ref{#1})}
\def\eqn#1{~(\ref{#1})}

\def\Cbf {{\bf C}}
\def\Ibf {{\bf I}}
\def\ybf {{\bf y}}
\def\xbf {{\bf x}}

\def\nbf {{\bf n}}
\def\albf{{\rm a}}
\def\xbft{{\bf x}^T}

\def\SS  {{\Sigma}}
\def\albfHat{\widehat{{\rm a}}} 
\def\da{\delta\albfHat}
\def\dT{\delta T}

\def\etal   {{\it et al.~}}

\def\microk {$\mu$K~}
\def\mmicrok{$\mu$K}
\def\microm {$\mu$m~}

\def\ie{{\frenchspacing\it i.e.}}

\def\ith{i^{th}}

\def\expec#1{\langle#1\rangle}

\def\spose#1{\hbox to 0pt{#1\hss}}
\def\simlt{\mathrel{\spose{\lower 3pt\hbox{$\mathchar"218$}}
    \raise 2.0pt\hbox{$\mathchar"13C$}}}
\def\simgt{\mathrel{\spose{\lower 3pt\hbox{$\mathchar"218$}}
    \raise 2.0pt\hbox{$\mathchar"13E$}}}
\def\simpropto{\mathrel{\spose{\lower 3pt\hbox{$\mathchar"218$}}
    \raise 2.0pt\hbox{$\propto$}}}

\def\Fig#1{Figure~\ref{#1}}

\def\rn{\noindent\parshape 2 0truecm 8.8truecm 0.3truecm 8.5truecm}
\def\nn#1 #2{#1, #2.}				% Name with 1 initial
\def\nnn#1 #2 #3{#1, #2. #3.}			% Name with 2 initials
\def\nnnn#1 #2 #3 #4{#1, #2. #3. #4.}		% Name with 3 initials
\def\nnnnn#1 #2 #3 #4 #5{#1, #2. #3. #4. #5.}	% Name with 4 initials
\def\rg#1;#2;#3;#4;#5;#6 {\par\rn#1 #2, {\it #3}, {\bf #4}, #5 (``#6'') \par}
\def\rf#1;#2;#3;#4;#5 {\par\rn#1 #2, {\it #3}, {\bf #4}, #5\par}
\def\rfbook#1;#2;#3;#4;#5 {{\frenchspacing\par\rn#1 #2, {\it #3} (#4: #5)\par}}
\def\rfproc#1;#2;#3;#4;#5;#6 {{\frenchspacing\par\rn#1 #2, in {\it #3}, ed. #4 (#5: #6)\par}}
\def\rfprep#1;#2;#3  {{\par\rn#1 #2, #3\par}}
\def\rfprepp#1;#2;#3 {{\par\rn#1 #2, #3\par}}

\journalid{337}{15 January 1989}
\articleid{11}{14}
\slugcomment{\today; Submitted to ApJ}
\begin{document}
\twocolumn[%%% Begin front material

\title{The Quest for Microwave Foreground X}

\author{
Ang\'elica de Oliveira-Costa\footnote{Department of Physics \& Astronomy, University of Pennsylvania, Philadelphia, PA 19104, USA}, 
                Max Tegmark$^{a}$,
%      Douglas P. Finkbeiner\footnote{Department of Astronomy, Princeton University, Princeton, NJ 08544, USA}, \\	     
                 R.D. Davies\footnote{University of Manchester, Jodrell Bank Observatory, Macclesfield, Cheshire, SK11 9DL, UK},\\
	 Carlos M. Gutierrez\footnote{Instituto de Astrofisica de Canarias, 38200 La Laguna, Tenerife, Spain},
	        A.N. Lasenby\footnote{Astrophysics Group, Cavendish Laboratory, Cambridge, CB3 0HE, UK}
                  R. Rebolo$^{c}$,
                R.A. Watson$^{b,c}$}

% \author{Ang\'elica de Oliveira-Costa\altaffilmark{1},
%                          Max Tegmark\altaffilmark{1}, 
%                Douglas P. Finkbeiner\altaffilmark{2}, \\                      
%                          R.D. Davies\altaffilmark{3},
%                  Carlos M. Gutierrez\altaffilmark{4}, 
%                         A.N. Lasenby\altaffilmark{5}, \\
%                            R. Rebolo\altaffilmark{4},
%                          R.A. Watson\altaffilmark{3,4}}
% 
% \altaffiltext{1}{Department of Physics \& Astronomy, University of Pennsylvania, Philadelphia, PA 19104, USA, angelica@higgs.hep.upenn.edu}
% \altaffiltext{2}{Department of Astronomy, Princeton University, Princeton, NJ 08544, USA}  
% \altaffiltext{3}{University of Manchester, Nuffield Radio Astronomy Laboratories, Jodrell Bank, Macclesfield, Cheshire, SK11 9DL, UK}
% \altaffiltext{4}{Instituto de Astrofisica de Canarias, 38200 La Laguna, Tenerife, Spain}
% \altaffiltext{5}{Mullard Radio Astronomy Observatory, Cavendish Laboratory, Madingley Road, Cambridge, CB3 0HE, UK}

%%%%%%%%%%%%%%%%%%%% ABSTRACT: %%%%%%%%%%%%%%%%%%%%%%%%%%%

\begin{abstract}
The WMAP team has produced a foreground map that can account for 
most of the low-frequency Galactic microwave emission in the WMAP 
maps, tentatively interpreting it as synchrotron emission.
Finkbeiner and collaborators  have challenged these conclusions, 
arguing that the WMAP team ``synchrotron'' template is in fact 
not dominated by synchrotron radiation, but by some dust-related 
Galactic emission process, perhaps spinning dust grains, 
making dramatically different predictions for its behavior
at lower frequencies.
By cross-correlating this ``synchrotron'' template  
with 10 and 15 GHz CMB observations, we find that its spectrum 
turns over in a manner consistent with spinning dust emission, 
falling about an order of magnitude below what the synchrotron 
interpretation would predict.
% "That's no ordinary rabbit."   
% "Go and tell your master that we have been charged by God with a sacred quest."  
% "We are the knights who say 'ni!'"  
%%
\end{abstract}

\keywords{cosmic microwave background  
-- diffuse radiation
-- radiation mechanisms: thermal and non-thermal
-- methods: data analysis}
]%%% End front material
  
%%%%%%%%%%%%%%%%%%% TEXT: %%%%%%%%%%%%%%%%%%%%%%%%%%%%%%%

\section{INTRODUCTION}

When doing precision cosmology with the Cosmic Microwave Background 
(CMB), a key challenge is accurately modeling and correcting for 
Galactic foreground contamination.
There are currently three indisputably identified Galactic foregrounds:
synchrotron radiation, free-free emission and thermal (vibrational) 
emission from dust grains. In the last few years, however, evidence
for the existence of a fourth component began to mount. This component, 
nicknamed  ``Foreground X'', is spatially correlated with 100 \microm 
dust emission but with a spectrum rising towards lower frequencies 
(in a manner that is incompatible with thermal dust emission), 
subsequently flattening and turning down somewhere around 15 GHz.

This fourth component was first discovered in the COBE DMR data by 
Kogut \etal (1996a, 1996b), who tentatively identified it as free-free 
emission\footnote{
	Draine and Lazarian (1998) argued against the free-free 
	hypothesis on energetic grounds, and suggested that 
	Foreground X was caused by dust after all, but via 
	rotational rather that vibrational emission.
	In the literature, Foreground X is sometimes called 
	``anomalous'' emission, and the most popular 
	interpretation has been that it is emitted by  
	spinning dust grains. 
	There is also the possibility that Foreground X is due to
	magnetic dipole emission from ferromagnetic grain materials
	(Draine and Lazarian 1999).}. 
It has since been detected in a variety of data sets such as
  Saskatoon (de Oliveira-Costa \etal 1997), 
  OVRO  (Leitch \etal 1997), 
  the 19 GHz survey (de Oliveira-Costa \etal 1998),
  Tenerife (de Oliveira-Costa \etal 1999, 2002; Mukherjee \etal 2001),
  QMAP (de Oliveira-Costa \etal 2000), 
  the ACME/SP94 data (Hamilton and Ganga, 2001), 
  Python V (Mukherjee \etal 2003),
  the WMAP data (Lagache 2003; Finkbeiner 2003), 
as well as in some other non-CMB data sets between 8-15 GHz 
(Finkbeiner \etal 2002; Finkbeiner \etal 2003, hereafter F03).
Adding pieces to this puzzle, joint re-analysis of COBE DMR 
\& the 19 GHz survey (Banday \etal 2003) and OVRO \& SP94 data 
(Mukherjee \etal 2002) also show the presence of Foreground-X.
Although a consensus about the existence of Foreground X had 
still not been reached, the case for spinning dust was 
beginning to look quite solid prior to the WMAP data 
release\footnote{
	{\it http://lambda.gsfc.nasa.gov/product/map/}.}. 

The WMAP team challenged this interpretation in their foreground 
analysis (Bennett \etal 2003b, hereafter B03b), arguing that there 
is still no solid evidence for an anomalous source of microwave 
emission at these wavelengths (\ie, no Foreground X).
They argued that all foreground emission observed at the five WMAP 
frequencies could be accounted for with a model involving only  
three foreground emission components, which they produced maps of and 
interpreted as (thermally radiating) dust, 
free-free emission and synchrotron radiation, respectively.

Using the Green Bank Plane Survey maps\footnote{
	The Green Bank Galactic Plane Survey maps the Galactic 
	plane within $\pm$4$^{\circ}$, with a resolution of 
	FWHM=11.2$^\prime$ at 8.35~GHz and FWHM=8$^\prime$ at 14.35~GHz. 
	See Langston \etal (2000) for more details.}, 
F03 have challenged these conclusions, arguing that the WMAP team 
``synchrotron'' template is in fact not dominated by synchrotron 
radiation, but by some dust-related Galactic emission process, 
perhaps spinning dust grains.
Given this dispute about whether the WMAP K-band 
``synchrotron map'' produced by B03b is really a map of synchrotron, 
let us hereafter refer to it as the {\it mystery map}, to avoid 
interpretative language. 

The logic behind identifying the WMAP team dust and free-free maps 
with these two physical emission sources is that they not only 
exhibit the right frequency dependence (by construction, from 
the way they were created) but spatially resembled external maps
of 100 \microm dust emission and H$\alpha$ emission, respectively.
F03 argue that, in constrast, the WMAP team mystery map does not 
strongly resemble external synchrotron maps at lower frequencies 
such as the 408 MHz Haslam map (Haslam \etal 1982).
%%   the 1.4 GHz Reich \& Reich map (Reich \& Reich 1986), 
%%   and the Rhodes 2.3 GHz map (Jonas \etal 1999). 
%%
B03b argue that this is due to strong spatial variations in the 
spectral index, whereas F03 argue that it is because the map 
(essentially defined a component having a synchrotron-like 
spectrum across the WMAP frequencies 23-93 GHz) is dominated 
by dust-related emission, not by synchrotron emission. 
 
The reason that such anomalous dust emission is so hard to pin
down is that its intensity is predicted to peak around 10-15 GHz, 
making it difficult to detect in the available radio surveys 
below 10 GHz such as those used by B03b (where it becomes 
subdominant to synchrotron radiation) and in the WMAP data at 
23GHz and above (where its spectral index is similar to synchrotron 
and free-free emission).
A smoking gun test for Foreground X therefore involves quantifying 
its behavior around 10-15 GHz, to see whether its spectrum continues 
to rise towards lower frequencies like synchrotron radiation or turns over.
 
It is important to point out, however, that although F03 used a data 
set in the desirable frequency range (8 \& 14 GHz) for the study of  
Foreground X, their survey was highly confined to the Galactic 
plane ($|b|<4^{\circ}$). In this context, one could argue that F03 
results do not necessarily extend to higher Galactic latitudes.
The purpose of this {\it Letter} is to investigate if such behavior can 
be extendible to higher Galactic latitudes and larger
areas of the sky. In order to do so, we use
the Tenerife 10 and 15 GHz maps, cross-correlating them with the 
WMAP K-band mystery map.

\begin{figure}[tb]
%\vskip-0.4cm
%\centerline{\hglue-5mm{\vbox{\epsfxsize=9.0cm\epsfysize=8.0cm\epsfbox{qaz_corr2_final.ps}}}}
%\vskip-2.0cm
\vskip-0.4cm
\centerline{\hglue-5mm{\vbox{\epsfxsize=9.0cm\epsfysize=8.0cm\epsfbox{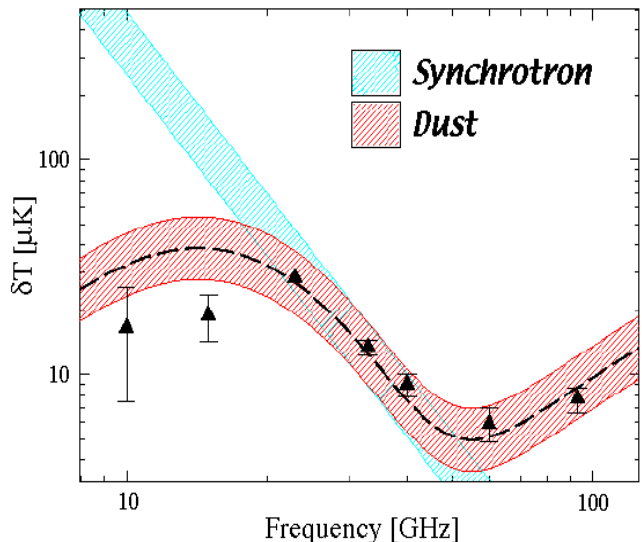}}}}
\vskip-0.5cm
\figcaption{\label{correlations}
    Contribution of the foreground traced by the mystery map 
    (the WMAP K-band ``synchrotron'' template) at different 
    frequencies for the $|b|>20^\circ$ Galactic cut (triangles).
%%%%%%%%%%%%%%%%%%%%%%%%%%%%%%%%%%%%%%%%%%%%%%%
%   Galactic cuts of 
%   $10^\circ$ (pentagons), 
%   $20^\circ$ (triangles) and
%   $30^\circ$ (squares).
%%%%%%%%%%%%%%%%%%%%%%%%%%%%%%%%%%%%%%%%%%%%%%%
    Shaded bands show models for synchrotron radiation 
    (cyan/light grey) and Foreground X emission (red/grey),
    assuming a synchrotron spectrum $\delta T\propto\nu^{-2.8}$ and
    a spinning dust spectrum from Draine \& Lazarian (1998),
    normalized to the $|b|>20^{\circ}$ cut at 23 GHz added to a 
    20K thermal dust component normalized at 90 GHz.
%   that best fit the $|b|>20^{\circ}$ cut results.
    }
\end{figure}

\begin{table}
{\footnotesize\center{\label{CorrTab2} Table~1 -- Correlations with the mystery map.}
\begin {center}
\begin {tabular}{|l|l|rrr|}	 
\hline
 \multicolumn{1}{|c|}{$|b|$}	             &
 \multicolumn{1}{ c|}{Map}                   &
 \multicolumn{1}{ r }{$\albfHat\pm\da$}      & 
 \multicolumn{1}{ r }{${\albfHat\over\da}$}  &	 
%\multicolumn{1}{ c }{$\sigma_{Gal}$}        &	       
 \multicolumn{1}{ c|}{$\dT^{(a)}$}          \\
 \multicolumn{1}{|c|}{}	   &
 \multicolumn{1}{ c|}{}    &
 \multicolumn{1}{ r }{}    & 
 \multicolumn{1}{ r }{}    &	 
% \multicolumn{1}{ c }{[\mmicrok]}&	       
 \multicolumn{1}{ c|}{[\mmicrok]}\\
\hline 
%%%%%%%%%%%%%%%%%%%%%%%%%%%%%%%%%%%%%%%%%%%%%%%%%%%%%%%%%%%%%%%%%%%%%%%%%%%%%%%%%%%%
% WITH POINT SOUCES:
%%%%%%%%%%%%%%%%%%%%%%%%%%%%%%%%%%%%%%%%%%%%%%%%%%%%%%%%%%%%%%%%%%%%%%%%%%%%%%%%%%%%
%$> 20^{\circ}$  &Ten10GHz  &-0.54 $\pm$0.55  &     -0.9   &21.1  &-11.3$\pm$11.6 \\
%		 &Ten15GHz  & 0.69 $\pm$0.23  &{\bf  3.0}  &	  & 14.5$\pm$ 4.8 \\
%%%%%%%%%%%%%%%%%%%%%%%%%%%%%%%%%%%%%%%%%%%%%%%%%%%%%%%%%%%%%%%%%%%%%%%%%%%%%%%%%%%%
%$> 20^{\circ}$  &Ten10GHz  &-0.35 $\pm$0.36  &     -0.9   &21.1  & -7.4$\pm$ 7.6 \\
%		 &Ten15GHz  & 0.61 $\pm$0.20  &{\bf  3.0}  &	  & 12.9$\pm$ 4.2 \\
%		 &WMAP-K    & 1.49 $\pm$0.07  &{\bf 21.3}  &	  & 31.3$\pm$ 1.5 \\
%		 &WMAP-Ka   & 0.72 $\pm$0.05  &{\bf 14.4}  &	  & 15.1$\pm$ 1.1 \\
%		 &WMAP-Q    & 0.51 $\pm$0.05  &{\bf 10.2}  &	  & 10.7$\pm$ 1.1 \\
%		 &WMAP-V    & 0.33 $\pm$0.05  &{\bf  6.6}  &	  &  6.9$\pm$ 1.1 \\
%		 &WMAP-W    & 0.37 $\pm$0.05  &{\bf  7.4}  &	  &  7.8$\pm$ 1.1 \\
%%%%%%%%%%%%%%%%%%%%%%%%%%%%%%%%%%%%%%%%%%%%%%%%%%%%%%%%%%%%%%%%%%%%%%%%%%%%%%%%%%%%
% WITHOUT POINT SOUCES:
%%%%%%%%%%%%%%%%%%%%%%%%%%%%%%%%%%%%%%%%%%%%%%%%%%%%%%%%%%%%%%%%%%%%%%%%%%%%%%%%%%%%
 $> 20^{\circ}$  &Ten10GHz  & 0.88 $\pm$0.48  &      1.8   & 16.5$\pm$ 8.9 \\
		 &Ten15GHz  & 1.05 $\pm$0.26  &{\bf  4.0}  & 19.6$\pm$ 4.9 \\
		 &WMAP-K    & 1.55 $\pm$0.07  &{\bf 22.1}  & 28.9$\pm$ 1.3 \\
		 &WMAP-Ka   & 0.73 $\pm$0.06  &{\bf 12.2}  & 13.7$\pm$ 1.1 \\
		 &WMAP-Q    & 0.49 $\pm$0.06  &{\bf  8.2}  &\ 9.2$\pm$ 1.1 \\
		 &WMAP-V    & 0.31 $\pm$0.06  &{\bf  5.2}  &\ 5.8$\pm$ 1.1 \\
		 &WMAP-W    & 0.41 $\pm$0.05  &{\bf  8.2}  &  7.7$\pm$ 0.94\\
%
% $> 20^{\circ}$  &Ten10GHz  & 0.88 $\pm$0.48  &      1.8  &18.7  & 16.5$\pm$ 8.9 \\
%		 &Ten15GHz  & 1.05 $\pm$0.26  &{\bf  4.0}  &	  & 19.6$\pm$ 4.9 \\
%		 &WMAP-K    & 1.55 $\pm$0.07  &{\bf 22.1}  &      & 28.9$\pm$ 1.3 \\
%		 &WMAP-Ka   & 0.73 $\pm$0.06  &{\bf 12.2}  &      & 13.7$\pm$ 1.1 \\
%		 &WMAP-Q    & 0.49 $\pm$0.06  &{\bf  8.2}  &      &\ 9.2$\pm$ 1.1 \\
%		 &WMAP-V    & 0.31 $\pm$0.06  &{\bf  5.2}  &      &\ 5.8$\pm$ 1.1 \\
%		 &WMAP-W    & 0.41 $\pm$0.05  &{\bf  8.2}  &      &  7.7$\pm$ 0.94\\
%%%%%%%%%%%%%%%%%%%%%%%%%%%%%%%%%%%%%%%%%%%%%%%%%%%%%%%%%%%%%%%%%%%%%%%%%%%%%%%%%%%%
\hline
\end{tabular}
\end{center}
} 
%%%%%%%%%%%%%%%%%%%%%%%%%%%%%%%%%%%%%%%%%%%%%%%%%%
\vskip-0.1cm
{\small $^{(a)}$ $\dT \equiv (\albfHat\pm\da) \sigma_{Gal}$, $\sigma_{Gal}=18.7\mu$K}\\
\end{table}

\section{CROSS-CORRELATION METHOD}\label{sec2}

To clarify the Foreground X puzzle, our goal is to measure the 
frequency dependence of the emission traced by the WMAP mystery map
down to 10 GHz.
We do this by cross-correlating it with the Tenerife observations 
(Guti\'errez \etal 2000, hereafter G00), which remain the most accurate 
large-area sky maps at 10 and 15 GHz and have not previously been 
spatially compared with the WMAP maps. 
Details of the cross-correlation method can be found in 
de Oliveira-Costa \etal (1997). This method models the vector of 
the Tenerife data $\ybf$ as a sum
\beq{signals}
	\ybf = \xbf\albf + \xbf_{CMB} + \nbf,
\eeq
where $\nbf$ is a vector that contains the Tenerife detector 
noise, $\xbf$ is a vector that contains the foreground template 
convolved with the Tenerife triple-beam (\ie, $\xbf_{i}$ would 
be the $\ith$ observation if the sky had looked like the mystery 
map), and ``$\albf$'' is a number that gives the level at which this 
foreground template is present in the Tenerife data.
 
The estimate of ``$\albf$'' is computed by minimizing 
    $\chi^2 \equiv
    (\ybf - \xbf\albf)^T \Cbf^{-1} (\ybf - \xbf\albf)$, 
where the covariance matrix $\Cbf$ includes both the 
experimental noise $\nbf$ and CMB sample variance 
in the CMB signal $\xbf_{CMB}$. Note that these quoted 
error bars include the effects of both noise and chance 
alignments between the Tenerife data and the template map.
The minimum-variance estimate of ``$\albf$'' is
\beq{alpha}
   \albfHat = {\xbft \Cbf^{-1} \ybf\over\xbft \Cbf^{-1} \xbf},
\eeq
with variance
\beq{varalpha}
   {\SS} \equiv \expec{\albfHat \albfHat^T} - 
                \expec{\albfHat} 
		\expec{\albfHat^T} =
         \left[ \xbft \Cbf^{-1} \xbf \right]^{-1}.
\eeq

We also perform correlations between the mystery map and 
less noisy data sets such as the five WMAP CMB maps. 
In these cases, the noise variance is so small that its 
exact value is irrelevant for our particular applications, 
so we simply use a diagonal covariance matrix 
	$\Cbf = \sigma^2 \Ibf$.
Equations \eqn{alpha} and \eqn{varalpha} then 
become simply
\beq{alpha2}
   \albfHat = {\xbf\cdot\ybf\over\xbf\cdot\xbf}, \quad  
   {\SS} = {\sigma^2\over |\xbf|^2}.
\eeq
Note that when using \eqq{varalpha}, we evaluate $\Cbf$ using the known 
noise properties of the Tenerife data (G00).
However, when using \eqq{alpha2}, we simply measure $\sigma$ directly 
from the data, since it corresponds to the scatter around the
best fit linear fit: 
\beq{scatter}
   \sigma = \frac{1}{N}  \sum_{i=1}^{N} 
            \left( y_i - \albfHat\xbf \right)^{1/2}.
\eeq

We use the latest version of the Tenerife data (Mukherjee \etal 2001), 
which has the estimated point source contribution (G00) removed 
before calculating the correlations. We also used the first-year WMAP 
CMB maps at 23 (K-band), 33 (Ka-band), 40 (Q-band), 60 (V-band) \& 
93 (W-band) GHz (Bennett \etal 2003a), the WMAP K-band mystery 
map\footnote{
	We also removed point sources from the mystery map
 	before calculating the correlations. Using the point sources catalogue 
	of B03b, we replaced the temperature for all pixels 
	centered within a circle of 1$^{\circ}$ radius around each 
	source by the mean temperature in an annulus
	between 1$^{\circ}$ and 2$^{\circ}$ around the source. 
	This removal does not have a major effect on the results. 
%	as a cross-check, we also performed 
%	correlations using the mystery map without point source removal 
%	and after setting the pixels within 1$^{\circ}$ equal to zero rather than
%	equal to the surrounding mean, neither of which changed the results 
%	substantially. 
	Removing the point sources lowers the rms fluctuations in the
	mystery map by about 10\%.}
 (B03b), and a foreground-cleaned CMB map made from a combination 
of the five WMAP CMB maps (Tegmark \etal 2003, hereafter TOH03). 
Before performing the correlations, all maps were convolved 
with the Tenerife triple-beam function, and data not overlapping 
the Tenerife observing region was discarded.  

\section{Cross-correlation results}

Cross-correlation results are presented in \Fig{correlations}
and Table 1. All fits were done for one template 
at a time, and statistically significant  ($>2\sigma$) 
correlations listed in this table are in boldface. 
%%
%Since the fluctuation levels depend strongly on Galactic 
%latitude, we perform our analysis for three different 
%latitude cuts: $10^{\circ}$, $20^{\circ}$ and $30^{\circ}$.
%By using these traditional cuts, we are able to 
%compare our results with others in the literature.
We perform our basic analysis with a $|b|>20^{\circ}$
Galactic latitude cut to facilitate comparison with 
older results in the literature.

\subsection{Validation of the Tenerife maps against WMAP}

As a first reality check, we cross-correlate the Tenerife data 
with the TOH03 map for a $|b|>20^{\circ}$ Galactic cut, obtaining 
$\albfHat=0.80\pm 0.12$ at 15 GHz and $\albfHat=0.59\pm 0.19$ 
at 10 GHz. This provides a beautiful validation of the Tenerife 
observations, showing a detection of CMB signal at about the 
$7\sigma$  level at 15 GHz and at the  $3\sigma$ level at 10 GHz.
This also confirms and further strengthens the results of 
Lineweaver \etal (1995), who cross-correlated Tenerife and 
COBE DMR data (Smoot \etal 1992; Bennett \etal 1996) at high 
Galactic latitudes and concluded that the CMB signal was 
consistent between the two data sets.
Using only the smaller sky area outside a $|b|>30^{\circ}$ 
Galactic cut, we still detect the CMB signal at about the 
$6\sigma$ level at 15 GHz but no longer obtain a significant 
detection at 10 GHz, where the Tenerife noise levels are 
higher.

This cross-correlation with WMAP allows an independent test of 
the calibration of the Tenerife measurements. For a $|b|>20^{\circ}$ 
Galactic cut, the above-mentioned numbers show that the relative 
calibration between WMAP and Tenerife is marginally consistent 
with being unity, with a slight preference for lower Tenerife values
at the level of $1.7\sigma$ at 15 GHz and $2.2\sigma$ at 10 GHz.
Down-calibrating Tenerife by say 10\% would be comfortably consistent 
with both the WMAP correlations reported here and previous estimates 
by the Tenerife team.
The 10 GHz error bar should be taken with a grain of salt, since 
the 10 GHz Tenerife map appears dominated by foregrounds rather 
than CMB. This means that the above-mentioned 10 GHz error 
bars are likely to underestimate the true uncertainty by not 
including the effect of chance foreground alignments with the 
TOH03 map.

In what follows, we down-calibrate the Tenerife maps by 10\% 
to be pedantic. Note that this has no effect whatsoever on 
our conclusions, which hinge on a discrepancy of an order of magnitude
rather than at this level of tens of percent.

\subsection{Foreground X results}

We then cross-correlate the Tenerife data with the WMAP-K mystery
map -- see \Fig{correlations} and Table~1 for results.
To extend \Fig{correlations} to higher frequencies,
we also perform the cross-correlation replacing the Tenerife data by 
the five WMAP CMB maps.
All measurements $\delta T$ plotted in \Fig{correlations} are 
obtained by multiplying the measured coefficient $\albfHat$ 
by $\sigma_{Gal}$, the rms fluctuations in the WMAP-K mystery 
map convolved with the Tenerife 
triple-beam. This means that one can interpret these numbers as being the 
signal that is explained by (or traced by) the WMAP-K mystery map 
at different frequencies.
%%
%% The squares, triangles and pentagons correspond to the correlations
%% at $30^{\circ}$,  $20^{\circ}$ and $10^{\circ}$, respectively. Upper 
%% limits are 2-$\sigma$. 
%% 
The triangles correspond to the correlations at $|b|>20^{\circ}$ cut. 
For comparison, we plot theoretical models for synchrotron 
(cyan/light grey) and Foreground X (or spinning dust -- red/grey) models,
assuming a synchrotron spectrum $\delta T\propto\nu^{-2.8}$ and
a spinning dust spectrum from Draine \& Lazarian (1998),
normalized to the $|b|>20^{\circ}$ cut at 23 GHz added to a 20K thermal dust 
component normalized at 90 GHz.

\Fig{correlations} shows that if the WMAP-K mystery map were 
strongly dominated by synchrotron emission, as assumed in B03b, 
then our $|b|>20^\circ$ $\delta T$-measurement should have
been about an order of magnitude larger at 15 GHz and about 
two order of magnitudes larger at 10 GHz. Even with a shallow
spectral slope $\nu^{-2.3}$, the discrepancy is a factor of five at 15 GHz. 
The fact that we observe a plateau instead, with a hint of a 
down-turn, strongly disfavors this synchrotron hypothesis 
and agrees better with the F03 interpretation that the mystery map 
is dominated by Foreground X, \ie, anomalous dust emission.

For this conclusion to be valid, we need to close an important 
loophole. B03b point out that the poor correlation between the 
mystery map and low-frequency synchrotron maps could be due to 
variations in the synchrotron spectral index across the sky.
Tegmark (1998) quantifies this effect, showing that if 
$\delta T\propto\nu^\alpha$ and the spectral index $\alpha$ varies 
with an rms $\Delta\alpha$ across the sky, then correlations 
of the type that we have measured are suppressed by a factor  
\beq{rEq}
	r\approx 
	e^{-{1\over 2}\left(\Delta\alpha\ln{\nu_2\over\nu_1}\right)^2},
\eeq
where $\nu_1$ and $\nu_2$ are the two frequencies involved.
Comparing $\nu_1=15$ GHz and $\nu_2=23 GHz$ for our $|b|>20^\circ$ 
case, an $\alpha=-2.8$ synchrotron model would predict 
	$\delta T(\nu_1)/\delta T(\nu_2)=(15/23)^{-2.8}\approx 3.3$
whereas we measured 
	$\delta T(\nu_1)/\delta T(\nu_2)=0.46\pm 0.15$, 
\ie, a factor of seven discrepancy. The MID foreground model from 
Tegmark {\etal} (2000) has $\Delta\alpha=0.15$, which according to 
\eq{rEq} gives a completely negligible suppression factor 
$r\approx 0.998$ (since the two frequencies are so close together).
Even the extreme assumption $\Delta\alpha=1$ gives the negligible 
suppression $r\approx 0.9$, \ie, a mere 10\% effect and nowhere 
near explaining the factor of seven.
In summary, our analysis is rather immune to the effect of spectral 
index variations, since the Tenerife observing frequencies are so 
close to those of WMAP. In contrast, \eq{rEq} shows that the correlation 
between the WMAP K-band and say the 408 MHz Haslam map would be suppressed 
by a dramatic factor of four for $\Delta\alpha=0.4$, since
the frequency lever arm is much longer.

Also, assuming that all rms signal seen by WMAP-K mystery map at 
$|b|>20^{\circ}$ cut is due to synchrotron emission, we should 
expect to have a synchrotron signal of
% 21.1(10/23)$^{-2.8} \approx$ 217.3 \microk at 10 GHz and 
% 21.1(15/23)$^{-2.8} \approx$  69.8 \microk at 15 GHz. 
  18.7(10/23)$^{-2.8}$ $\approx$ 192.6 \microk at 10 GHz and 
  18.7(15/23)$^{-2.8}$ $\approx$  61.9 \microk at 15 GHz. 
For comparison, the observed Tenerife rms CMB signal is 
  $\delta T_{\ell} = 30^{+15}_{-11}$ \microk (G00).
%%%%%%%%%%%%%%%%%%%%%%%%%%%%%%%%%%%%%%%%%%%%%%%%%%%%%%%%%%%%%%%%%%%%%%%%
% For comparison, the observed Tenerife rms signal at $|b|>20^{\circ}$ 
% cut is 
%   $\approx$ 168.9 \microk at 10 GHz and
%   $\approx$  27.3 \microk at 15 GHz,
% which is lower than the expected rms signal seen by WMAP-K 
% mystery map scaled down to the Tenerife frequencies.
%%%%%%%%%%%%%%%%%%%%%%%%%%%%%%%%%%%%%%%%%%%%%%%%%%%%%%%%%%%%%%%%%%%%%%%%
% RMS NOISE IN THE SLICE: 300.6\microk (10GHz) and 128.5\microk (15GHz). 
% WHAT IS Q_rms_ps = sqrt(rms**2) - (noise**2)? 
%%%%%%%%%%%%%%%%%%%%%%%%%%%%%%%%%%%%%%%%%%%%%%%%%%%%%%%%%%%%%%%%%%%%%%%%

\section{Discussion}

The spectacular WMAP data has provided crucial new information
about Foreground X. Although the WMAP analyses by Lagache (2003) 
and Finkbeiner (2003) suggest the presence of Foreground X across 
the WMAP frequencies (23-93 GHz), they cannot alone rule out the 
hypothesis of B03b that there is no Foreground X, merely a 
dust-correlated synchrotron component with strong spatial variations 
in the spectral index. 
The smoking gun test for Foreground X therefore involves quantifying 
its behavior around 10-15 GHz, to see whether its spectrum continues 
to rise towards lower frequencies like synchrotron radiation or 
turns over.

A joint analysis of the WMAP data with the Green Bank Galactic Plane 
Survey makes clear that the Foreground X provides a much better fit 
to the data between 8 and 93 GHz than the synchrotron hypothesis 
could (F03). It is important to bear in mind, however, that the Green 
Bank Galactic Plane Survey maps the sky only at very low Galactic 
latitudes, leaving open the question of whether the F03 results 
extend to the higher Galactic latitudes that are relevant for 
precision cosmology.
The Tenerife--WMAP-K mystery map correlations that we have measured 
provide the missing information in this puzzle, quantifying the behavior 
of Foreground X at higher Galactic latitudes than studied by F03 and 
lower frequencies than observed by WMAP.
 
Cross-correlating Tenerife with a foreground cleaned WMAP CMB map 
shows that the Tenerife CMB data agrees well with WMAP, detecting 
the WMAP CMB signal at about $7\sigma$.
Cross-correlating Tenerife with the WMAP-K mystery map shows that
the mystery map cannot be dominated by synchrotron emission at 
Galactic latitudes $|b|>20^\circ$, since this would give 15 GHz 
measurements almost an order of magnitude larger than observed 
and even worse discrepancies at 10 GHz.
Instead, we find that the Foreground X spectrum has a plateau 
and a hint of a down-turn below 20 GHz, just as predicted in 
spinning dust models.

\bigskip
\smallskip

We would like to thank Douglas Finkbeiner for discussions that 
motivated this work as well as for helpful comments on the manuscript.
We also would like to thank Bruce Draine for helpful comments on the 
manuscript.
Support was provided by the NASA grant NAG5-11099,
NSF grant AST-0134999, and fellowships from the David and Lucile
Packard Foundation and the Cottrell Foundation.  
We acknowledge the NASA office of Space Sciences, the WMAP 
flight team, and all those who helped to process and analyze
the WMAP data.

%%%%%%%%%%%%%%%%%%%%%% REFERENCES: %%%%%%%%%%%%%%%%%%%%%%%%%

\clearpage

\end{document}